# Manipulating exchange bias using all-optical helicity-dependent switching

P. Vallobra[1], T. Fache[1], Y. Xu[1], L. Zhang[1], G. Malinowski[1], M. Hehn[1], J.-C. Rojas-Sánchez[1] E.E. Fullerton[2] and S. Mangin[1*]

[1]Institut Jean Lamour (UMR 7198), Université de Lorraine, Vandœuvre-lès-Nancy, France
2 Center for Memory and Recording Research, University of California San Diego, La Jolla, CA 92093-0401, USA

* Correspondence to Stephane Mangin ( Stephane.mangin@univ-lorraine.fr)

Deterministic all-optical control of magnetization without an applied magnetic field has been reported for different materials such as ferrimagnetic and ferromagnetic thin films and granular recording media. These findings have challenged the understanding of all-optical helicity-dependent switching of magnetization and opened many potential applications for future magnetic information, memory and storage technologies. Here we demonstrate optical control of an antiferromagnetic layer through the exchange bias interaction using the helicity of a femtosecond pulsed laser on IrMn/[Co/Pt]$_{xN}$ antiferromagnetic/ ferromagnetic heterostructures. We show controlled switching of the sign of the exchange bias field without any applied field, only by changing the helicity of the light, and quantify the influence of the laser fluence and the number of light pulses on the exchange bias control. We also present the combined effect of laser pulses and applied magnetic field. This study opens applications in spintronic devices where the exchange bias phenomenon is routinely used to fix the magnetization orientation of a magnetic layer in one direction.

Since Louis Néel's prediction of an antiferromagnetic (AFM) spin ordering in 1936 [1] and the demonstration of AFM order in MnO by Henri Bizette in 1938 [2], antiferromagnetism has attracted increasing attention because of fascinating physics and major role in important and emerging applications for magnetic data storage, memories, sensors and logic devices [3]. Most of the current functionality of AFM materials arises from the exchange bias phenomenon first observed by Meiklejohn and Bean [4] on fine particles of cobalt with a cobalt-oxide shell. Exchange bias is often observed in AFM/ferromagnetic (FM) heterostructures as a field shift in the magnetization curve of the FM layer characterized by the exchange bias field ($H_{EB}$). This property has been extensively used in spintronic devices such as magnetoresistive heads [5] and magnetic random access memories [6].

In current devices AFM materials are used to control and stabilize the magnetization direction of FM layers. However, AFM materials are increasingly being considered for new applications in spintronics [7-10]. This interest arises in part because AFM materials are insensitive to magnetic fields, have high intrinsic resonant frequencies in the THz regime and because of the new possibility of probing and manipulating AFM layers using spin currents. For instance it has been shown that an AFM domain wall can be moved by spin transfer torque (STT) [11] [12] [3] and AFM order can be switched [3] [13].

The conventional approach to establish exchange bias field consists of heating the AFM/FM bilayer structure above the blocking temperature ($T_B$), close but lower than the Néel Temperature ($T_N$), the temperature at which the AFM spin lattice orders. The bilayer is then cooled from $T_B$ under a magnetic field sufficient to saturate the FM layer. The orientation of the FM layer magnetization sets the orientation of the interface AFM spins because of the interface exchange coupling between the two layers. Other ways of controlling and manipulating exchange bias has been very recently investigated such as using ionic motion at interfaces [14] or modifying crystal structure [13].

The fundamental mechanism explaining exchange bias has been discussed extensively [15-18] but remains a topic of continued interest [19] [14]. Ultrafast optical excitations of a AFM/FM exchange bias bilayer has been one approach to probe the interfacial interaction and interesting fast magnetization dynamics have already been observed [20-22]. A number of reports have found that photoexcitation of the AFM/FM interface induces large modulations in the exchange bias field on ultra-short time scales leading to coherent magnetization precession in the FM layers. Detailed time-resolved studies of dynamics determined that the characteristic time scale of laser-induced exchange-bias quenching in a polycrystalline Co/IrMn bilayer is 0.7±0.5 ps [22]. The fast decrease in exchange coupling upon laser heating is attributed to a spin disorder at the interface created by laser heating. In the present paper we demonstrate that exciting an exchange biased system using circularly polarized ultra-fast short laser pulses we can deterministically control the sign of the exchange bias without an applied magnetic field and control its amplitude.

For this study, we have grown well-known IrMn/[Co/Pt]$_{xN}$ exchange biased multilayer shown schematically in Fig. 1 for N=1. The choice was driven to design samples, which combine a large perpendicular exchange bias field [23] [24] with the possibility to exhibit all-optical helicity-dependent switching (AO-HDS) [25-29]. In [Co/Pt]x$_N$ multilayers with perpendicular anisotropy and low number of repeats N, it has been demonstrated that the magnetization can be switched deterministically by sweeping a femtosecond laser beam [28] or using a static beam [30]. Indeed it was previously shown that for [Co(0.6nm)/[Pt(2nm)]x$_N$ with N≤2 the criteria related to stable domain size needed to observe AO-HDS is fulfilled [31]. Note that for AOHDS with either a sweeping beam or a static beam, multiple laser pulses are needed to reverse magnetization deterministically [32].

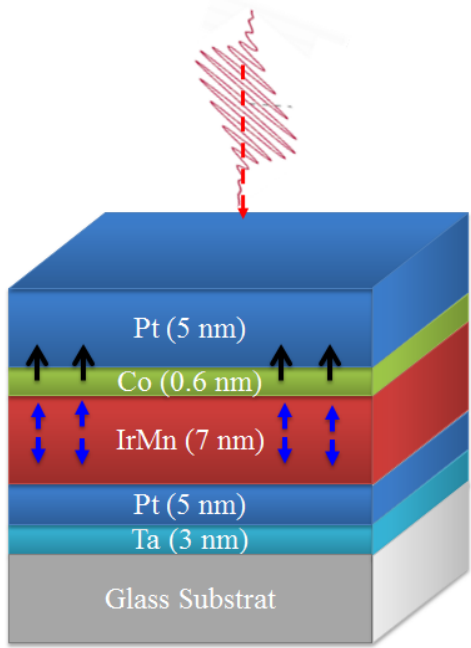

Figure 1: Schematic of the Interplay between femtosecond pulses and exchange bias for an IrMn(7nm)/[Co(0.6nm)/[Pt(2nm)] multilayers showing perpendicular exchange bias.

In exchange biased systems the hysteresis loop (magnetization vs. applied magnetic field H) of the FM layer is characterized by its width, the coercive field ($H_C$), and its horizontal shift along the field axis ($H_{EB}$) as shown in Figs. 2 and 3. The IrMn thickness (7 nm) has been optimized to observe a reasonable exchange bias effect and two samples with N=2 (Fig. 2) and N=1 (Fig. 3) have been chosen to carry out the study of the effect of polarized femtosecond pulses on the magnetic configuration of the bilayer.

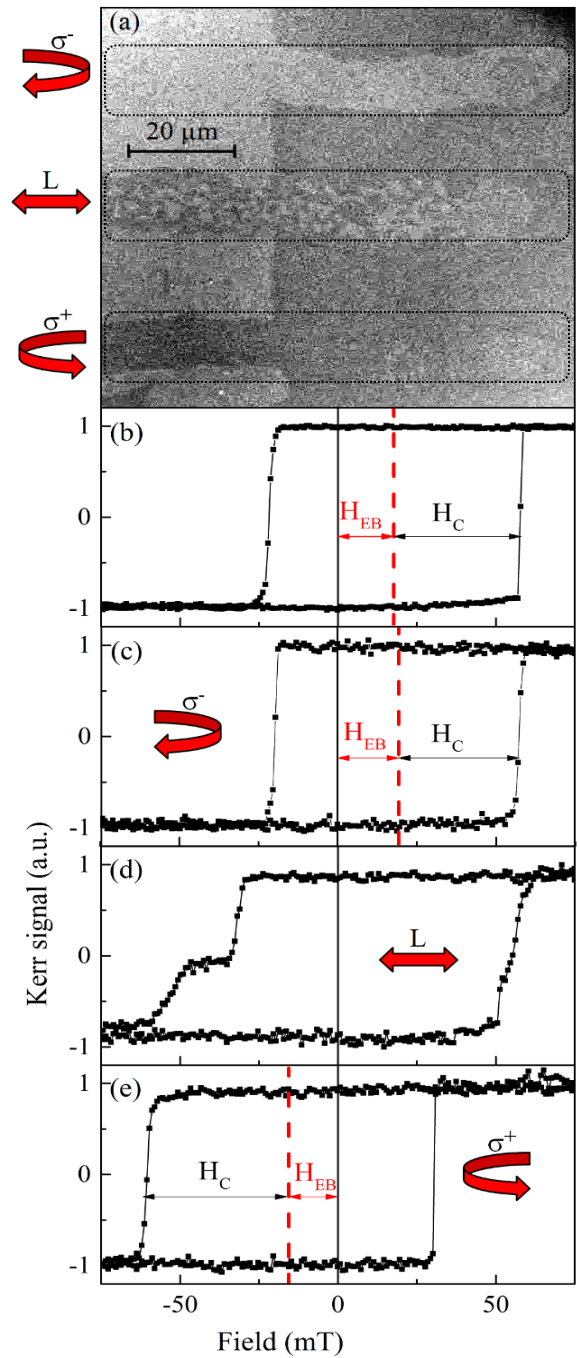

Figure 2): Results obtained on an exchange bias sample Glass/Ta(3nm)/Pt(5nm)/IrMn(7nm)/[Co(0,6nm)/Pt(2nm)]x2 /Pt(5nm) a) Faraday imaging after that, a right-circularly (σ⁻), a linear L and a left-circularly (σ⁺) polarized laser beam have been swept over the sample from right to left with a sweeping speed of approximately 10 μm/sKerr signal hysteresis loop obtained on . b) a non exposed sample area (as grown sample) c) on an area where a right-circularly polarised, (σ⁻) laser beam was swept, d) area where a linearly polarised (L) laser beam was swept e) area where a left-circularly polarised, (σ⁺) laser beam was swept,

Figure 2.b (resp figure 3.b) shows the small exchange bias effect observed for the as grown sample

with N= 2 (resp. N=1). For N=2 $H_C$= 38.2 mT and $H_{EB}$= 19.1 mT and for N=1 $H_C$= 34.9 mT and $H_{EB}$= -11.1 mT.

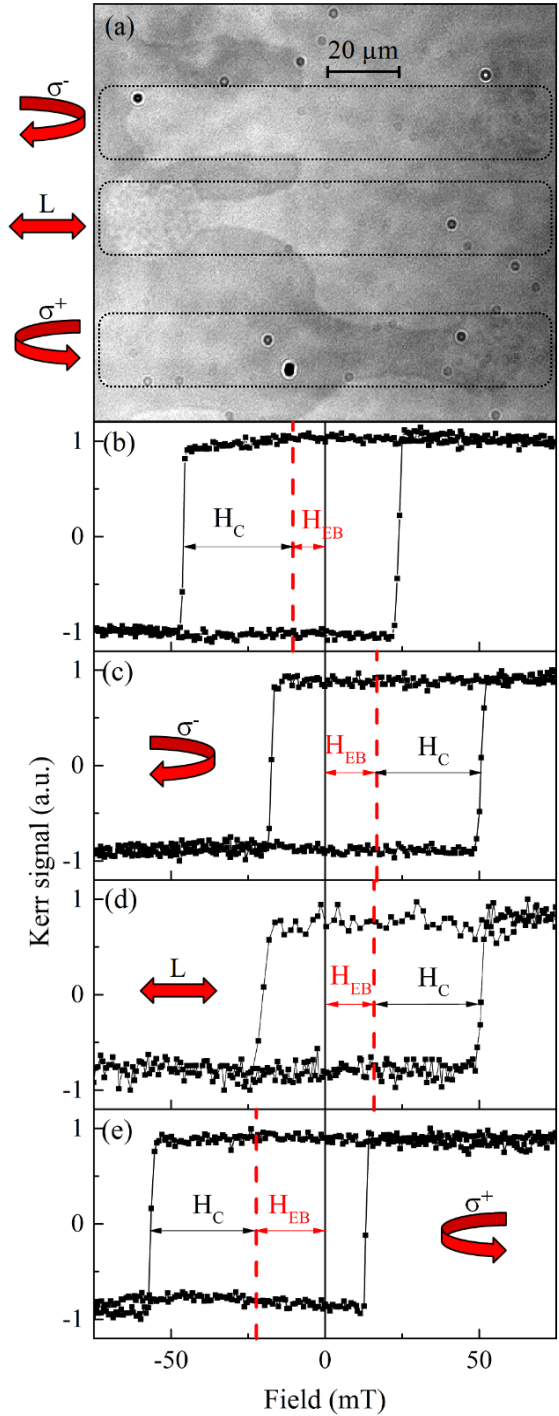

*Figure 3) : Results obtained on an exchange bias sample Glass/Ta(3nm)/Pt(5nm)/IrMn(7nm)/[Co(0,6nm)/Pt(2nm)]x1 /Pt(5nm) a) Faraday imaging after that, a right-circularly ($\sigma^-$), a linear L and a left-circularly ($\sigma^+$) polarized laser beam have been swept over the sample with a sweeping speed of approximately 10 µm/sKerr signal Hysteresis loop obtained on . b) a non exposed sample area (as grown sample) c) on an area where a right-circularly polarised, ($\sigma^-$) laser beam was swept, d) area where a linearly polarised (L) laser beam was swept e) area where a left-circularly polarised, ($\sigma^+$) laser beam was swept,*

**AO-HDS reversal and control of exchange bias**

In a first step, the laser has been shinned onto the Pt capping layer (Fig. 1) and the ability to optically control the orientation of the FM layer is check when exchange biased with the AFM layer. .Figures 2(a) and 3(a) demonstrate the AO-HDS effect on AFM/FM bilayer with N=2 and N=1, respectively. The final state of the magnetization of the sample is deterministically controlled by the helicity of the circularly polarized light when the laser beam is swept on the sample similarly to what was observed for Co/Pt multilayers [28]. For N=2 the linear helicity (L) creates multi-domain states of both orientations independent of the initial magnetic state. For N=1 such multi-domain state is not observed for linear helicity (L). In that case the thermal gradient produced by the laser light drags the domain wall as it has been previously seen [33]. These results demonstrate that AO-HDS of the [Co/Pt] multilayer is observed even with the additional exchange coupling due to the IrMn (AFM) layer.

In addition to the control of the FM magnetization orientation, the sign of the exchange bias is deterministically controlled by the AO-HDS process. The hysteresis loops reported in figures 2c-e and 3c-e were measured on areas where the laser beam was swept with $\sigma^-$, L and $\sigma^+$ polarization respectively. The comparison of Figs. 2.c to 2.e and Figs 3.c to 3.e shows that $\sigma^+$ polarized light induces a negative exchange bias while $\sigma^-$ polarized light induces a positive exchange bias. After sweeping, the bias direction is consistent with the final state of the $[Co/Pt]_{xN}$ magnetization, independent of the initial bias direction. The process is reversible and the sign of the exchange bias for the same area of the sample can be reversed with subsequent sweeping of the beam with opposite helicity. For the case of linearly polarized light the final loop depends on the sample (N=2 or N=1). In the case of linear polarization which results in a multidomain state configuration, the superposition of hysteresis loops with both positive and negative bias is observed [34]. However,when domain dragging is observed, the hysteresis loop is similar to the one obtained for $\sigma^-$ (see fig 3(d)). In all cases the exchange bias effect is set in agreement with the FM domain orientation set by AO-HDS process.

**Static beam experiments**

Static beam experiments with fluences comparable to those used in Figs. 2 and 3 have been performed by varying the number of pulses from 1 to 50000. Starting from the as grown sample, $\sigma^-$ helicity was used and local hysteresis loops were measured at different positions inside and outside the diskleft by the laser spot (Fig. 4). Since the laser intensity is larger in the center of the spot and decreases as one moves away from the center it allows us to study the influence of the laser intensity and number of pulses on the induced exchange bias. Figure 4 reports the evolution of the exchange bias (4.a) and the coercive field (4.b) as a function of the distance from the circle center after 1 pulse and 50000 pulses.

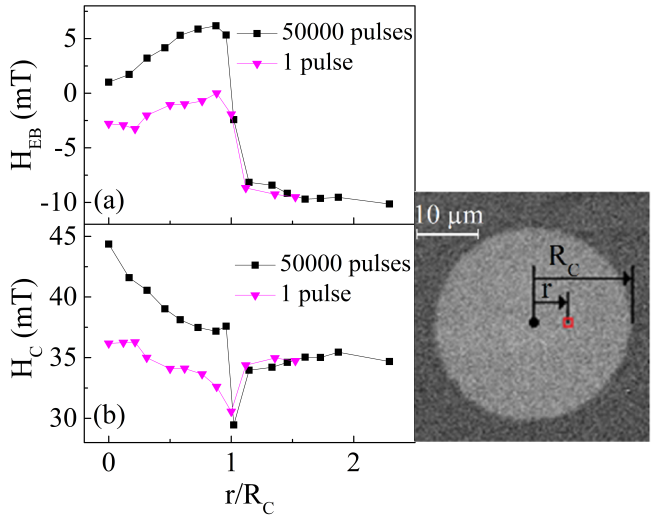

Figure 4: Exchange bias field ($H_{EB}$) and coercive field ($H_C$) as a function of the position (r) from the center of a laser spot after a single pule (full triangle) and after 50 000 pulses.

First the results show that for a radius (r) larger than a critical value ($R_C$) the effect of light on the exchange bias and the coercivity is negligible. In other words there is a critical intensity ($I_C$) below which the exchange bias state is unaffected by the optical excitation. The largest exchange bias field change is observed for an intensity just above $I_C$ where r is slightly smaller than $R_C$. For a single pulse the exchange bias field is significantly reduced while for a large number of pulses (50,000) the sign of the exchange bias field is deterministically changed from negative to positive for $\sigma^-$ helicity. These measurements demonstrate that multiple pulses are needed to control the sign of the exchange bias Field. This is consistent with earlier measurements showing that AO-HDS in Co/Pt multilayers requires multiple pulses [32].

**Static beam versus sweeping beam experiments**

While magnetic switching is observed in both static and sweeping beam experiments, the exchange bias field values measured after the static multiple pulse process are smaller than those obtained by sweeping the laser beam over the sample. This result could be explained by considering the different characteristic temperatures of the AFM/FM system and the time needed to cross those temperatures.

From the highest temperatures towards room temperature, the AFM/FM system has to cross first the Curie temperature ($T_C$) of the Co/Pt multilayer found to be above 650 K for the considered Co and Pt thicknesses. From previous measurements [28] it appears that another temperature $T_{SW}$ slightly smaller than $T_C$ could be defined as the temperature at which the light has the largest effect on magnetization switching of the FM layer. Finally, at lower temperatures, the IrMn AFM layer orders with a Neel temperature of $T_N$= 650 K and a blocking temperature of $T_B$= 550 K . For our AFM/FM system $T_C$, $T_{SW}$, $T_N$ and $T_B$ are close.

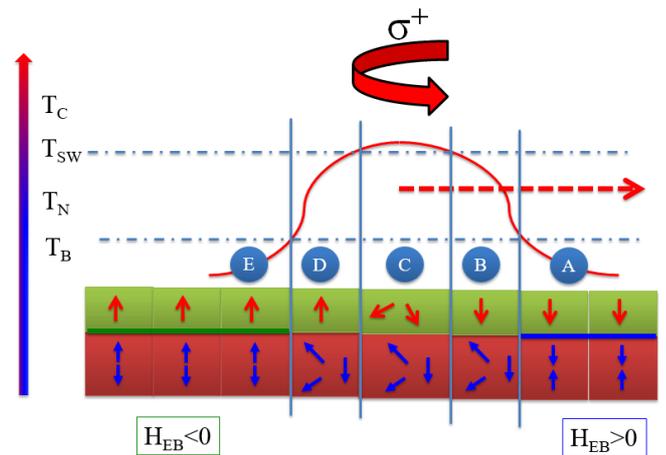

Figure 5: Sketch of the effect of a sweeping beam on the AFM/FM magnetic configuration In A) the laser power is too low to influence either the AFM and FM magnetic configuration. Initially, theFM magnetization is pointing down and the exchange bias is positive. For B) the power is large enough to demagnetized the AFM C) the left-circularly polarisation ($\sigma^+$) and the laser fluence allow the FM switching from down to up while the temperature is above the AFM Blocking temperature. D) the AFM layer is cooled down through its blocking temperature with a FM layer magnetization pointing up which leads to a negative exchange bias

Starting with a FM magnetization pointing down and a positive exchange bias, the laser is swept towards the right with a polarization that reverses the magnetization of the FM layer. As sketch in figure 5 In region A) the laser fluence is too small and the temperature is lower than $T_B$. The laser has no impact on the magnetic configuration, a situation that is similar to $r>R_C$ in Fig. 4. In region B) the temperature is larger than $T_B$. The AFM layer starts to be disoriented whereas the FM layer remains unaffected. In region C) the laser fluence and helicity is sufficient to reach $T_{SW}$ and to reverse the Co/Pt magnetization. In regions D) to E) the bilayer cools down through $T_B$ with a reversed magnetized FM layer which re-orients the AFM layer and so the exchange bias. Thus the orientation of the AFM is not set directly by the helicity of the light but by the orientation of the FM while crossing the blocking temperature.

The differences between the static and sweeping beam response could be due to the fact that $T_C$, $T_{SW}$, $T_N$ and $T_B$ are close. Indeed During sweeping, the FM layer has time to thermalized and saturated when the IrMn crosses $T_N$ and $T_B$ leading to a high exchange bias field. On the contrary, in the static experiment, when IrMn crosses $T_N$ and $T_B$, the FM layer is still fluctuating leading to a reduced exchange bias.

**Influence of FM orientation on setting the exchange bias**

Finally, to confirm the role of the orientation of the FM layer and the cooling process, we studied the combined effect of the laser and an applied magnetic field. Figure 6 compares the effect of sweeping a circularly polarized laser beam (Fig. 6.a) with a linearly polarized laser beam while a magnetic field sufficiently large to saturate the magnetization was applied (Fig. 6.b).

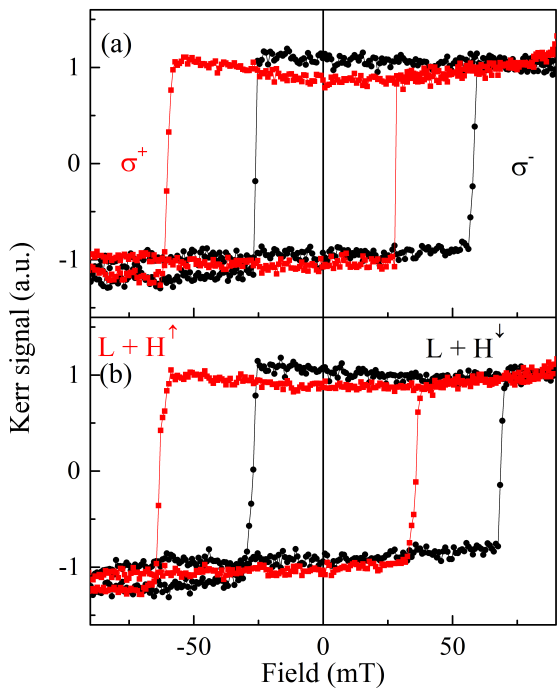

*Figure 6: Hysteresis loop of Glass/Ta(3nm)/Pt(5nm)/IrMn(7nm)/[Co(0,6nm)/Pt(2nm)]x1 /Pt(5nm) Comparison between a) the effect of a right-circularly ($\sigma^-$), or a left-circularly ($\sigma^+$) polarized laser beam with the combined effects of a linearly polarized light and an applied magnetic field, up or down.*

This demonstrates that the effect of a $\sigma^+$ (resp. $\sigma^-$) polarization on the exchange bias field and coercive field is similar to the effect of a linear polarization and a positive (H↑) (resp. negative(H↓)) applied magnetic field.

In the second experiment, the laser is swept with a fluence such that it does not affect the orientation of the Co/Pt layer magnetization. The magnetization was either set up or down by an applied magnetic field and then the laser was scanned with a reduced laser intensity. As shown in Fig. 7 the resulting exchange bias is then set by the orientation of the Co/Pt magnetization before the light exposure and no significant influence of the helicity of the light is observed.

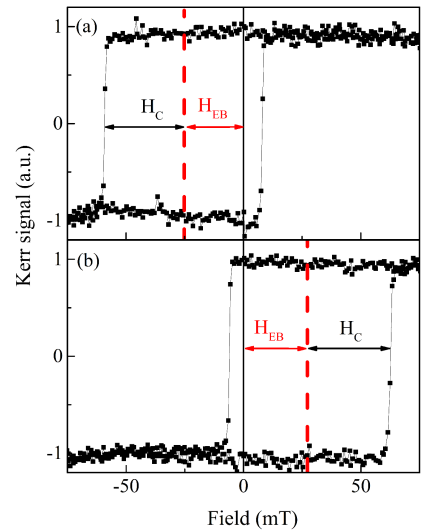

*Figure 7: Hysteresis loop of Glass/Ta(3nm)/Pt(5nm)/IrMn(7nm)/[Co(0,6nm)/Pt(2nm)]x1 /Pt(5nm) obtained in an area where a linearly polarized laser with a power just under the threshold has been swept beginning from a) saturation down and b) saturation up*

In that case the antiferromagnetic configuration is modified by the combined heating effects of the laser and the initial orientation of the Co/Pt layer.

In conclusion we have demonstrated that exchange bias can be manipulated with femtosecond laser pulses without any applied field. We showed that AFM magnetic configurations can be modified deterministically but indirectly with light only. The femtosecond polarized laser first modifies the FM orientation which set the exchange bias while cooling through the blocking temperature. The next perspectives consist to determine if the AFM layer can be directly manipulated without the help of a FM layer. In terms of applications our work may have strong impact on magnetic memories, logic and recording technology. The possibility to manipulate the exchange bias coupling locally as it was already suggested for the thermally-activated switching MRAM concept [Pre 07] would offer scalability, thermal stability, energy efficiency, low response to residual field.

Methods: To perform optical excitation, we use a Ti:sapphire fs-laser with a 5-kHz repetition rate, a wavelength of 800nm (1.55 eV), and pulse duration of 35 fs. The Gaussian beam spot is focused with a FWHM of approximately 50 μm and swept with a velocity of about 5 um/s and a fluence of $10^9 \, mJ/cm^2$ The response of the magnetic film was studied using a static Faraday microscope to image the magnetic domains while the laser is illuminating the sample. The helicity of the beam is controlled by a zero-order quarter-wave plate, which transforms linearly polarized light (L) into circularly left- (σ +) or right-polarized light (σ −). The present measurements are performed at room temperature and the laser beam was swept at a constant rate of 3–20μm. s−1 with the typical laser spot size of 50μm. Magnetic images were obtained in transmission using a magneto-optic faraday microscope. Magnetic hysteresis measurements were performed using a magneto-optic Kerr effect setup. The background signal was subtracted from the intial hysteresis loops. All of the samples were grown by d.c. magnetron sputtering from elemental sources onto room-temperature glass substrates. Alloys were grown by co-sputtering where the source powers controlled the composition. During the growth the base pressure was below 5 10-8 Torr. Multilayers and heterostructures were formed by sequential deposition